\documentstyle[preprint,aps]{revtex}
\begin{document}
\draft
\preprint{CLNS 96/1403}
\title{Spin Symmetry Without Heavy Quarks:\\
Hyperon Form Factors In The Large $N_c$ Limit}
\author{Chi-Keung Chow}
\address{Newman Laboratory of Nuclear Studies, Cornell University, Ithaca, 
NY 14853.}
\date{\today}
\maketitle
\begin{abstract}
In the large $N_c$ limit, all hyperon decays involving the same quark diagram 
$Q\to Q'$ are described by a single weak form factor $\eta_{QQ'}(w)$.  
No assumption on the mass of $Q$ or $Q'$ is necessary, making our results 
applicable to both $b\to c$ and $c\to s$ transitions.  
This same form factor describes both $\Lambda_Q\to\Lambda_{Q'}$ and 
$\Sigma^{(*)}_Q\to\Sigma^{(*)}_{Q'}$ transitions.  
The (non-)commutativity between the heavy quark and the large $N_c$ limits 
is briefly discussed under the definite example of $\Lambda_b\to\Lambda_c$ 
decay.  
\end{abstract}
\pacs{}
\narrowtext
In the past few years, our understanding in the large $N_c$ limit of QCD in 
the baryon sector has been greatly improved.  
It is now realized that a contracted SU(4) spin-flavor symmetry arises in 
the large $N_c$ limit, and many important consequences of this result have 
been discussed in Ref.~\cite{1}\footnote{Besides this work by the San Diego 
group based on the ``current algebra approach'', there are other groups 
studying the large $N_c$ limit of baryon dynamics in other formalisms, like 
the Harvard group and the Berkeley group.  
Our reference to the San Diego group results reflects the author's 
familiarity with that particular formalism and certainly does not imply 
that the works by other groups are inferior or irrelevant.  
The physics is independent of which formalism one works with, as the crux 
of the matter is the SU(4) spin-flavor symmetry, which is present in all 
different approaches.}.  
It turns out that baryon properties like the masses and the meson couplings 
are severely constrained by this spin-flavor symmetry.  
In this letter, the application of this symmetry to weak processes of 
hyperons are discussed.  
For hyperon we mean a baryon with a single $s$, $c$ and $b$ quark.  
It is found that the reduction of the numbers of form factors, a result 
usually obtained in the heavy quark limit, is reproduced, although no 
assumptions has been made on the masses of the quarks involved in the weak 
decay.  
As a result, our results are equally applicable to ``heavy-to-heavy'' (like 
$b\to c$) and ``heavy-to-light'' (like $c\to s$) decays.  
In the large $N_c$ limit, a single universal weak form factor $\eta_{QQ'}(w)$ 
describes both $\Lambda_Q\to\Lambda_{Q'}$ and $\Sigma^{(*)}_Q\to 
\Sigma^{(*)}_{Q'}$, where $Q$ and $Q'$ can be $b$, $c$ or $s$.  
Moreover, one can use flavor SU(3) to extend this result to decays to $u$ and 
$d$ quarks as well.  
This form factor $\eta_{QQ'}(w)$ in general depends on the quark species $Q$ 
and $Q'$ but is independent on the spin structure of the baryon and the 
current involved.  
However, $\eta_{QQ'}(w)$ is in general not normalized at any kinematic point 
unless we further assume some flavor symmetry between the parent and daughter 
quark.  

Since the baryons are heavy in the large $N_c$ limit, their velocities are 
well-defined.  
\begin{mathletters}
A weak transition from a spin-$1\over2$ hyperon of velocity $v$ to one 
also with spin $1\over2$ but velocity $v'$ is in general parametrized by six 
form factors.  
\begin{eqnarray}
&&\langle B'_{Q'}(v',s')|\bar {Q'}\gamma^\mu Q|B_Q(v,s)\rangle \nonumber\\
&&= \bar u(v',s')(F_1(w)\gamma^\mu + F_2(w)v^\mu + F_3(w)v'^\mu)u(v,s),
\end{eqnarray}
\begin{eqnarray}
&&\langle B'_{Q'}(v',s')|\bar {Q'}\gamma^\mu\gamma_5 Q|B_Q(v,s)\rangle 
\nonumber\\
&&= \bar u(v',s')(G_1(w)\gamma^\mu + G_2(w)v^\mu + G_3(w)v'^\mu)\gamma_5 
u(v,s),
\end{eqnarray}
where $B_Q$ can be a $\Lambda_Q$ or a $\Sigma_Q$ hyperon.  
\end{mathletters}
In general the set of form factors $F$'s and $G$'s governing the $\Lambda$ and 
$\Sigma$ sectors are unrelated.  
In additional to these twelve form factors (six for $\Lambda$ and six for 
$\Sigma$) there are another eight form factors describing the $\Sigma_Q\to 
\Sigma^*_{Q'}$ transition.  
So in general one needs twenty form factor to describe all hyperon weak 
transitions involving the quark level process $Q\to Q'$.  
(The $\Sigma^*_Q$ hyperons can decay electromagnetically to $\Sigma_Q$, so 
the $\Sigma^*_Q\to\Sigma^{(*)}_{Q'}$ transitions are physically 
insignificant.)  

Now it is well known that, in the heavy quark limit, the number of independent 
form factors is dramatically reduced \cite{2a,2b,2c,2d}.  
For example, for the decay $\Lambda_Q \to \Lambda_{Q'}$, there are only two 
independent form factors $\eta(w)$ and $\beta(w)$ when $m_Q\to\infty$, 
\begin{eqnarray}
F_1(w)=\eta(w)-\beta(w),\quad F_2=2\beta(w),\quad F_3(w)=0,\nonumber\\
G_1(w)=\eta(w)+\beta(w),\quad G_2=2\beta(w),\quad G_3(w)=0.  
\label{ff}
\end{eqnarray}
Moreover, when we further assume $m_{Q'}\to\infty$, $\beta(w)$ vanishes 
identically and the weak decay is described by just a single universal form 
factor $\eta(w)$.  
The drastic simplification bases on the observation that, in the heavy quark 
limit, the heavy quark spin $s_Q$ is conserved, i.e., 
\begin{equation}
[s_Q,H_Q] = 0, 
\label{spin}
\end{equation}
where $H_Q$ is the hamiltonian of a baryon containing the heavy quark $Q$.  
Similar simplification is also possible for the $\Sigma_Q\to\Sigma_{Q'}$ 
transition, which is described by ten form factors when $m_Q\to\infty$ but 
only two when $m_{Q'}$ is also set to infinity.  
It must be emphasized that this reduction of form factors is just the 
consequence of the spin symmetry (\ref{spin}).  
As a result, this reduction of form factors will also occur at other limits 
of QCD where condition (\ref{spin}) is satisfied.  
We will see promptly that this is indeed the case for baryons in the large 
$N_c$ limit.  

The representations of the baryonic states in the large $N_c$ limit has been 
studied in detail in Ref. \cite{1}.  
It is found that the baryons states can be denoted by $|I,I_3;J,J_3;K\rangle$,
where $\vec K=\vec I+ \vec J$ is an additional operator satisfying the SU(2) 
commutation relations.  
We have $K=0$ for baryons with just $u$ and $d$ quarks, while a hyperon with 
a single $s$, $c$ or $b$ quark has $K={1\over2}$.  
In the latter case $K$ generates the rotation of the $s$, $c$ or $b$ quark 
in question, i.e., $\vec K=\vec s_Q$.  
Under this notation, the baryon hamiltonian in the large $N_c$ limit has
the following expansion in orders of $1/N_c$: 
\begin{equation}
H=N_c M_0 + N_c^0 s_Q M_1 + N_c^{-1} (a I^2 + b J^2 + c s_Q^2).  
\end{equation}
Since for hyperons with a single $s$, $c$ or $b$ quark we have $s_Q 
\equiv 1/2$, the first two terms are just constant numbers which commutes 
with $\vec I$, $\vec J$ and $\vec K$.  
The $\Sigma_Q-\Lambda_Q$ and $\Sigma^*_Q-\Sigma_Q$ splittings are determined 
by the parameters $a$ and $b$ respectively.  
It is evident that, in the large $N_c$ limit, $\Lambda_Q$ and $\Sigma^{(*)}_Q$ 
are degenerate, and the splittings enter only at order $N_c^{-1}$, two orders 
below the leading term.  
Hence, if we only keep the first two terms in the expansion, the hamiltonian 
is just a constant number (total degeneracy between all states) and condition 
(\ref{spin}) is satisfied.  
As a result, we have ``heavy quark spin symmetry'' although we have not placed 
any assumptions on the mass of the ``heavy quark'', which may as well be just 
a strange quark.  
This symmetry is simply the consequence of the {\it light quark} spin-flavor 
symmetry in the large $N_c$ limit.  
Roughly speaking, this large $N_c$ light quark spin symmetry means that the 
physics is invariant upon the flipping of the spin of any light quark.  
As a result, one can flip the spin of the heavy quark by first flipping the 
spins of all the light quarks, and then rotation the whole system by an angle 
$\pi$.  
The physics is invariant under both processes, and hence we have ``heavy quark 
spin symmetry'' without necessarily a heavy quark.  

So we have come to see that all consequences of the heavy quark spin symmetry 
are also valid in the large $N_c$ limit.  
For example, the heavy quark spin symmetry decrees that $\Sigma_Q$ and 
$\Sigma^*_Q$ are degenerate.  
Our study shows that they are also degenerate in the large $N_c$ limit, in 
regardless of the mass of the heavy quark.  
In particular, the $\Sigma^*-\Sigma$ splitting is of order $N_c^{-1}$.  
This is in accordance with the experimental result 
\begin{equation}
\Sigma^*-\Sigma \sim 200 \hbox{ MeV} \ll 400 \hbox{ MeV} \sim K^*-K, 
\end{equation}
as the $K^*-K$ splitting is not $1/N_c$ suppressed at all.  
Moreover, if one takes the unconfirmed measurement of $\Sigma^*_c$ mass at 
2530 MeV \cite{4a,4b}, one have a corresponding inequality in the charmed 
sector as well.  
\begin{equation}
\Sigma^*_c-\Sigma_c \sim 75 \hbox{ MeV} \ll 145 \hbox{ MeV} \sim D^*-D.  
\end{equation}

Returning to the weak transition, with the spin symmetry we have the reduction 
of weak form factors as usual.  
In the large $N_c$ limit, the $\Lambda_Q\to\Lambda_{Q'}$ transition is 
described by a single form factor $\eta(w)$.  
\begin{equation}
\langle \Lambda_{Q'}(v')|\bar{Q'}\Gamma Q|\Lambda_Q(v)\rangle 
=\eta_{QQ'}(w) \bar u_{\Lambda_{Q'}} \Gamma u_{\Lambda_Q}, 
\end{equation}
where the $\Sigma^{(*)}_{Q'}\to\Sigma^{(*)}_Q$ transitions are controlled by 
two form factors.  
\begin{eqnarray}
&&\langle\Sigma^{(*)}_{Q'} (v')|\bar {Q'} \Gamma Q |\Sigma^{(*)}_Q(v)
\rangle\nonumber\\ 
&&= ({\zeta_1}_{QQ'}(w) g_{\mu\nu} + {\zeta_2}_{QQ'}(w) v_\nu v'_\mu)\,
\bar u^\nu_{\Sigma^{(*)}_{Q'}}(v') \,\Gamma\, u^\mu_{\Sigma^{(*)}_Q}(v) , 
\end{eqnarray}
where $u^\nu_{\Sigma^*_Q}(v')$ is the Rarita--Schwinger spinor vector for a 
spin-$\textstyle {3\over2}$ particle and $u^\mu_{\Sigma_Q}(v,s)$ is defined by
\begin{equation}
u^\mu_{\Sigma_Q}(v) = {(\gamma^\mu + v^\mu)\gamma_5 \over \sqrt 3}
u_{\Sigma_Q}(v)
\end{equation}
and similarly for $u^\nu_{\Sigma^{(*)}_{Q'}}(v')$.  
Since we have not utilized the heavy quark limit, these reduction of form 
factors are equally applicable for ``heavy-to-heavy'' transitions like 
$\Lambda_b\to\Lambda_c$ as well as ``heavy-to-light'' ones like $\Lambda_c\to 
\Lambda$.  
The latter case is particularly interesting, as in this case the large $N_c$ 
reduction of former factors is more powerful than that in the $N_c=3$ heavy 
quark limit.  
(One form factor in the former case, in contrast to two for the latter.)  
In the notation of Eq.~(\ref{ff}), we have $\beta(w)=0$ in the large $N_c$ 
limit.  
More will be said about the interpretation of this statement.  

Further reduction of form factors is still possible.  
In Ref. \cite{5}, it has been proved that in the large $N_c$ limit, the 
weak form factors in the $\Lambda_Q$ and $\Sigma_Q$ sectors are related.  
(This result first appeared in Ref. \cite{6} in the context of the chiral 
soliton model.)  
In the notation above, we get 
\begin{equation}
{\zeta_1}_{QQ'}(w) = -(1+w){\zeta_2}_{QQ'}(w) = \eta_{QQ'}(w) .  
\label{univ}
\end{equation}
Hence we have reached the main result of this article: {\it in the leading 
order of the large $N_c$ limit, all baryon transitions involving the same 
quark level diagram $Q\to Q'$ are described by a single form factor, in 
regardless of the masses of $Q$ and $Q'$.}  
It should not be a surprising result.  
It is well known that, in the large $N_c$ limit, the static properties of 
the tower states are closely related.  
They all have the same mass in the first two orders in the $1/N_c$ expansion, 
and their axial current couplings are simply interrelated by Clebsch--Gordan 
coefficients.  
Our study just shows that such interrelations also hold for weak form factors. 

On the other hand, some important implications in the heavy quark limit 
cannot be reproduced here.  
Here we do not have the heavy quark {\it flavor} symmetry, and the 
$\eta_{QQ'}(w)$ for different $Q$ and $Q'$ are in general unrelated.  
Moreover, the weak form factor in the large $N_c$ limit is in general not 
normalized.  
Of course, if one in addition assume some flavor symmetry between the initial 
and final quarks $Q$ and $Q'$, the form factor will be normalized at certain 
kinematic points.  
For example, when both $Q$ and $Q'$ are heavy, the form factor is normalized 
at the point of zero recoil $w=1$.  
Unfortunately, for a ``heavy-to-light'' decay, no flavor symmetry is 
applicable and the form factor is not normalized throughout the whole 
kinematic range.  

Several points of discussion are in place here.  
Firstly, we have restricted ourselves to weak transitions between $s$, $c$ 
and $b$ quarks.  
It is natural to ask if our results can be applied to weak transitions 
producing $u$ and $d$ quarks, like $c\to d$ or $b\to u$.  
The answer is affirmative.  
Under SU(3) flavor symmetry, the same form factor describes, say, $c\to s$ 
and $c\to d$ transitions.  
So the applicability of our results to the former case warrants that to the 
latter.  
So our results are in fact not only valid for hyperons but all (orbitally 
unexcited) baryons.  

Another point of interest is the asymptotic behavior of these form factors 
is the large $N_c$ limit.  
For $w\neq 1$, i.e., when the initial and final baryons have different 
velocities, the weak transition involves changing the momenta of all the 
quarks inside the baryons\footnote{I would like to thanks M. Lu for a 
discussion on this point.}.  
As a result, one expect the transition amplitude to be highly suppressed 
when $N_c$ is large.
This expectation is indeed verified in the special cases of $\Lambda_b\to
\Lambda_c$ \cite{7} and $\Sigma^{(*)}_b\to\Sigma^{(*)}_c$ \cite{6}, where 
the form factors scales like $\exp(-N_c^{3/2})$ in the large $N_c$ limit.  
So, when we claim above in Eq.~(\ref{univ}) $\eta_{QQ'}(w)={\zeta_1}_{QQ'}(w)$,
that should be read as 
\begin{equation}
{{\zeta_1}_{QQ'}(w)\over\eta_{QQ'}(w)}=1+O(1/N_c), 
\end{equation}
with both form factors vanishing in the large $N_c$ limit when $w\neq 1$.  
Similarly, when we assert that for a ``heavy-to-light'' decay $\beta(w)$ 
vanishes in the large $N_c$ limit, it should be read as 
\begin{equation}
{\beta(w)\over\eta(w)}=O(1/N_c).  
\end{equation}

As mentioned before, the lack of any absolute normalization of the form 
factors may limit the usefulness of these results.  
Hence it is tempting to assume heavy quark symmetry from the outset and 
perform a double expansion of $1/N_c$ and $1/m_Q$.  
This, however, depends on the commutativity of the heavy quark and the 
large $N_c$ limit, which is a highly nontrivial assumption.  
It has been shown that the chiral limit and large $N_c$ limit does not always 
commute \cite{9a,9b,9c}, and the non-commutativity is embodied in this ratio:  
\begin{equation}
d={m_\pi\over m_\Delta - m_N}, 
\end{equation}
the numerator and the denominator measuring the deviations from the chiral 
and large $N_c$ limits respectively.  
So going to the chiral limit amounts to setting $d=0$, while in the large 
$N_c$ limit we have $d\to\infty$.  
Clearly these two conditions cannot be both satisfied in the same time.  
Experimentally $d\sim 0.5$, so the real world does not resemble any of these 
two limiting cases very closely.  
We will spend the rest of the letter discussing the similar non-commutativity 
issue between the heavy quark and the large $N_c$ limit, and use the leading 
order corrections to the $\Lambda_b\to\Lambda_c$ decay as our prime example.  

It has been proved that, when $m_b\to\infty$ and arbitrary $m_c$, the 
$\Lambda_b\to\Lambda_c$ decay is controlled by two form factors, which are 
denoted by $\eta(w)$ and $\beta(w)$ is Eq.~(\ref{ff}).  
Then one can expand these form factors in orders of $1/m_c$.  
\begin{mathletters}
\begin{equation}
\eta(w)=\eta_0(w)+\textstyle{1\over m_c}\eta_1(w)+\textstyle{1\over m_c^2} 
\eta_2(w) + \dots\;, 
\end{equation} 
\begin{equation}
\beta(w)=\beta_0(w)+\textstyle{1\over m_c}\eta_1(w)+\textstyle{1\over m_c^2} 
\beta_2(w) + \dots\;.  
\end{equation} 
\end{mathletters}
In leading order of $1/m_c$, i.e., when the $c$ quark is also heavy, we have 
$\beta_0(w)=0$.  
The leading term of $\beta(w)$ appears at the first order of $1/m_c$, which 
is calculated in Ref. \cite{8}.  
\begin{equation}
\beta_1(w)=-\eta_0(w)\delta(1+w)^{-1}, 
\end{equation}
where 
\begin{equation}
\delta={\bar\Lambda\over m_c} = {m_{\Lambda_c}-m_c\over m_c} .  
\end{equation}
Since the numerator $\bar\Lambda\sim N_c$, it seems that the first order 
correction diverges in the large $N_c$ limit, in contradiction of our claim 
above that $\beta(w)=0$ when $N_c$ is large.  

To get a better understanding of this paradox, one needs to retrace the 
derivation in Ref.~\cite{8}, be careful to keep all the terms of positive 
powers of $\delta$, which are legitimately discarded in Ref.~\cite{8} as 
$N_c=3$ in their work.  
The $1/m_c$ correction arises from the matching of current in full QCD 
$\bar c \Gamma b$ to that in the effective theory $\bar h_c \Gamma h_b$.  
The effective field $h_c$ is related to the quark field $c$ in full QCD by 
\begin{mathletters}
\begin{equation}
c=\exp(-im_c v'\cdot x)\left(1-{i/\!\!\!\!D\over m_c}\right)^{-1}h_c.  
\end{equation}
in the ``Harvard formulation'' of heavy quark effective theory, or 
alternatively 
\begin{equation}
c=\exp(-im_c v'\cdot x)\exp\left({i/\!\!\!\!D\over m_c}\right)h_c.  
\end{equation}
in the ``Mainz formulation''\footnote{See Ref. \cite{11a,11b} for a comparison 
between the two formulations of heavy quark effective theory.  
The two different formulations give identical predictions to physical 
quantities like scattering matrix elements, so the choice of formulation 
should be immaterial.}.  
\label{corr}
\end{mathletters}
When acting on a hadron state, the derivative operator in the effective 
theory measures the residual momentum, which is typically of the order of the 
mass of the light degrees of freedom.  
\begin{equation}
{i/\!\!\!\!D\over m_c}\sim {\bar\Lambda\over m_c}=\delta.  
\end{equation}
It is clear that the heavy quark expansion is an expansion in $\delta$.  
When $N_c=3$, $\delta$ is formally small and Eqs.~(\ref{corr}) reduces to the 
normal relation 
\begin{equation}
c=\exp(-im_c v'\cdot x)\left(1+{i/\!\!\!\!D\over m_c}\right)h_c.  
\end{equation}
When $N_c\to\infty$, however, the expansion scheme becomes ambiguous.  
The size of the correction term depends on how we approach the double limits 
of heavy quark and large $N_c$; the relative size of the two expansion 
parameters is described by the quantity $\delta$.  
In the heavy quark limit one have $\delta=0$, while with large $N_c$ we 
get $\delta\to\infty$.  
The situation is analogous to that in Ref.~\cite{9a,9b,9c}, where the 
non-commutativity between the chiral and large $N_c$ limits are characterized 
by the parameter $d$.  
Experimentally $\bar\Lambda\sim$ 700 MeV while $m_c\sim$ 1500 MeV, so 
$\delta\sim 0.5$ and the real world is not especially close to both limiting 
cases.  
Our studies suggested that the apparent divergence of $\beta(w)$ in the large 
$N_c$ limit is an artifact of the truncation to the first term in the $1/m_c$ 
expansion, while the whole $\beta(w)$, with the contributions from all orders 
of $1/m_c$ summed, should be at most of order $N_c^{-1}$.  

The discussion above shows that an $1/N_c$ expansion in the heavy quark 
limit is more subtle than one might naively expect.  
Since the heavy quark symmetry is more predictive than the large $N_c$ limit 
for ``heavy-to-heavy'' decays, and since $\delta<1$ suggested that the real 
world has more resemblance to the heavy quark limit than to the large $N_c$ 
limit, we do not expect that our results will be highly important to the 
understanding of these decays\footnote{One can certainly not exclude the 
possibility that a careful reformulation of the heavy quark effective 
theory may allow such a double expansion.  
Such pursuit, however, is beyond the scope of this letter.}.  
In our opinion, our results are best applied to ``heavy-to-light'' decays, 
where the large $N_c$ result is more predictive than the heavy quark 
counterparts.  
Moreover, it is known that nine different form factors, four local and five 
non-local, are needed to parametrize the $1/m_Q$ corrections to the 
``heavy-to-light'' decay form factors \cite{10}.  
Since only six form factors are necessary in the most general formulation, 
it means that the heavy quark limit loses all predictive power for such 
transitions at order $1/m_Q$.  
Therefore it is interesting to study the $1/N_c$ corrections for such decays.  
If our scheme can retain its predictive power at $1/N_c$, it will be a 
practical alternative expansion scheme to study the ``heavy-to-light'' or 
even ``light-to-light'' transitions.  
So we conclude by stating that the large $N_c$ limit simplifies weak 
transition matrix elements dramatically, but the $1/N_c$ corrections are not 
well understood yet and deserves further investigation.  

\acknowledgements
I am grateful to Ming Lu, Dan Pirjol and Tung--Mow Yan for discussions.  
This work is supported in part by the National Science Foundation.

\end{document}